\documentclass{elsart}

\usepackage{graphicx}
\usepackage{amssymb}

\begin{document}
\begin{frontmatter}




\title{CPT Violation and the Nature of Neutrinos}


\author[a1]{G. Barenboim}\ead{gabriela@fnal.gov},
\author[a2]{J.F. Beacom}\ead{beacom@fnal.gov},
\author[a3]{L. Borissov}\ead{liubo@phys.columbia.edu},
\author[a1]{B. Kayser}\ead{boris@fnal.gov}

\address[a1]{Theoretical Physics Department, Fermi National 
Accelerator Laboratory, Batavia, Illinois 60510-0500, USA}
\address[a2]{NASA/Fermilab Astrophysics Center, Fermi National 
Accelerator Laboratory, Batavia, Illinois 60510-0500, USA}
\address[a3]{Physics Department, Columbia University, New York, 
New York 10027, USA}

\begin{abstract}
In order to accommodate the neutrino oscillation signals from the
solar, atmospheric, and LSND data, a sterile fourth neutrino is
generally invoked, though the fits to the data are becoming more and
more constrained.  However, it has recently been shown that the data
can be explained with only three neutrinos, if one invokes CPT
violation to allow different masses and mixing angles for neutrinos
and antineutrinos.  We explore the nature of neutrinos in such
CPT-violating scenarios.  Majorana neutrino masses are allowed, but in
general, there are no longer Majorana neutrinos in the conventional
sense.  However, CPT-violating models still have interesting
consequences for neutrinoless double beta decay.  Compared to the
usual case, while the larger mass scale (from LSND) may appear, a
greater degree of suppression can also occur.
\end{abstract}


\end{frontmatter}

\vspace{-0.5cm}
{\small FERMILAB-Pub-02/014-T, FERMILAB-Pub-02/014-A}
\vspace{-0.5cm}


\newcommand{\be}{\begin{equation}}
\newcommand{\ee}{\end{equation}}
\newcommand{\bea}{\begin{eqnarray}}
\newcommand{\eea}{\end{eqnarray}}
\def\p{{\bf p}}


\section{Introduction}

In recent years, stronger and stronger experimental evidence for
neutrino oscillations has been accumulating.  As is well-known, this
evidence would extend the Standard Model by requiring neutrino masses
and mixings.  While knowing the values of the mass and mixing
parameters may be an important clue to physics beyond the Standard
Model, more information is needed.  For example, it is presently
unknown whether neutrino processes violate lepton number or not, or
whether the neutrinos are their own antiparticles or not.

Neutrino masses and mixings can be straightforwardly included in the
Standard Model, as can the choice of Majorana or Dirac neutrinos.  But
in fact, the present neutrino oscillation evidence may imply more
profound extensions to the Standard Model.  With three neutrinos,
there are only two independent mass-squared differences.  However,
since the mass-squared differences required by the solar\cite{solar},
atmospheric~\cite{atm}, and LSND~\cite{lsnd} neutrinos are $\delta m^2
\simeq$ $10^{-5}$ eV$^2$, $10^{-3}$ eV$^2$, and $1$ eV$^2$,
respectively, three different mass-squared scales are needed to
explain all the data.  The usual way out is to postulate a fourth
neutrino, thus allowing three independent mass-squared differences.
This fourth neutrino must be sterile under the weak interactions since
the invisible width of the $Z^\circ$ only allows three light active
neutrinos.

The LSND results are based on an appearance experiment, namely
$\bar{\nu}_\mu \rightarrow \bar{\nu}_e$.  For years, the solar and
atmospheric neutrino results only gave clear evidence for the
disappearance of $\nu_e$ or $\nu_\mu, \bar{\nu}_\mu$, respectively.
Thus one or the other could have been accommodated by oscillations to
sterile neutrinos.  However, the difficulties that seem to require the
introduction of a fourth neutrino have recently become more acute as
the combined solar neutrino results from the Sudbury Neutrino
Observatory and Super-Kamiokande indicate that $\nu_e \rightarrow
\nu_\mu, \nu_\tau$, and the atmospheric neutrino results from
Super-Kamiokande indicate that $\nu_\mu,\bar{\nu}_\mu \rightarrow
\nu_\tau,\bar{\nu}_\tau$.  Thus with three signals of neutrino
oscillations among active flavors, there is not only a problem of not
enough independent mass-squared differences, but also a problem of
where to incorporate the required mixing with the sterile neutrino.
While four-neutrino models may still work, it is only with difficulty
(see, e.g., Ref.~\cite{concha}).

Recently, an intriguing but speculative suggestion to accommodate
these results has been made~\cite{cpt1,cpt2} (see also
Ref.~\cite{murayama}).  If CPT is violated in the neutrino sector,
then the mixing parameters which govern solar neutrino oscillations of
$\nu_e \rightarrow \nu_\mu, \nu_\tau$ do not have to be the same as
those that govern $\bar{\nu}_\mu \rightarrow \bar{\nu}_e$ in LSND.
Without introducing new particles or interactions to the Standard
Model, this allows the introduction of a new mass-squared difference
and it also solves the problem of which flavors are oscillating to
which.  In the simplest model, atmospheric neutrino oscillations of
$\nu_\mu \rightarrow \nu_\tau$ behave the same as their antiparticles.

Thus, it is not merely that the Standard Model must be extended to
accomodate neutrino mass and mixing.  With the present data, taking
all of the experimental results at face value, one can either
introduce new particles (sterile neutrinos), or allow CPT violation.
Either suggestion, if confirmed, would be an important clue for
understanding physics well beyond the Standard Model.  Large CPT
violation in the neutrino sector, while a radical suggestion, should
be easily testable in the upcoming experiments.

Previous papers have explored the consequences of large CPT violation
in the neutrino
sector~\cite{cpt1,cpt2,murayama,skadhauge,bilenky,strumia,bahcall,greenberg,mocioiu}.
Here we explore the consequences for the Majorana or Dirac nature of
the neutrinos (for an introduction in the CPT-conserving case, see
Ref.~\cite{kayser}).  For CPT-conserving neutrinos, it is frequently
stated that ``If neutrinoless double beta decay is observed, then
neutrinos are Majorana particles,'' meaning that a neutrino and its
antineutrino are the same state.  We show that for CPT-violating
neutrinos, a more careful statement must be made, i.e., ``If
neutrinoless double beta decay is observed, then there must be
Majorana neutrino mass terms,'' and that in general, neutrinos are
{\it not} equal to the corresponding antineutrinos as states.  In
either the CPT-conserving or CPT-violating case, the observation of
neutrinoless double beta decay would of course imply the violation of
lepton number.  Needless to say, if one can determine the nature of
the neutrino masses, light will be shed on the symmetries of
fundamental particle interactions.

We also show that there are interesting practical consequences for
neutrinoless double beta decay, a process that violates lepton number
by two units.  One way to explain the present null\footnote{Though
evidence for neutrinoless double beta decay has very recently been
claimed in Ref.~\cite{KK}, it has been disputed in
Refs.~\cite{strumia,aalseth}.}  results~\cite{2breviews} in the usual
CPT-conserving scenario would be to say that the neutrino masses are
all below say, 0.1 eV (larger masses can be accommodated if the mixing
angles and Majorana phases cause cancellations).  Given present
constraints on the oscillation parameters, there can be a non-zero
minimum allowed effective mass.  In the CPT-violating model the
overall scale may be set by the large LSND mass scale of $\sim 1$ eV,
i.e., predicting a larger signal.  However, a larger degree of
suppression can occur, so that the effective mass can also be
arbitrarily small.


\section{CPT Violation and the Neutrino Masses}

In the Standard Model, extended to include neutrino masses, the
interactions are lepton-number (L) conserving.  Any nonconservation of
L would come from Majorana mass terms, which turn a neutrino into an
antineutrino.  {\it When CPT is conserved} and Majorana mass terms are
present, the neutrino mass eigenstates $\nu_i$ are Majorana particles.
That is, each $\nu_i$ is its own antiparticle in the sense that
\be
CPT \left| \nu_i \right\rangle = e^{i\xi_i} \left| \nu_i \right\rangle\,,
\ee
where $\xi_i$ is a phase.  Through the process pictured in Fig.~1,
exchange of the Majorana $\nu_i$ leads to neutrinoless double beta
decay ($\beta\beta_{0\nu}$).  This is the L-nonconserving reaction in
which one nucleus decays to another with the emission of two
electrons.  Conversely, even if $\beta\beta_{0\nu}$ should arise
predominantly from some mechanism other than light neutrino exchange,
the observation of this decay would imply that the electron neutrino
$\nu_e$ has a nonzero Majorana mass.  When CPT is conserved, this
would in turn imply that the mass eigenstates $\nu_i$ are Majorana
particles.  In a CPT-conserving world, when there are no
Majorana mass terms, L is conserved, $\beta\beta_{0\nu}$ is forbidden,
and the mass eigenstates $\nu_i$ are Dirac particles.  That is, each
$\nu_i$ differs from its CPT conjugate by the value of the conserved
quantum number L: $\nu_i$ has ${\rm L\ } = +1$ while $\bar{\nu}_i$ has
${\rm L\ } = -1$.

To see what becomes of this picture {\it when CPT is violated}, we
consider the simplest case of a single neutrino $\nu$ (i.e., $\nu_e$),
coupled to the electron by the Standard Model weak coupling, and its
CPT conjugate $\bar{\nu}$, coupled to the positron.  When allowing CPT
violation, we do not assume that $\nu$ and $\bar{\nu}$ have the same
mass.  Rather, we suppose that for a given spin direction, the
$\nu,\bar{\nu}$ mass matrix $M_\nu$ has the form
\be
M_\nu = \left[\begin{array}{cc}
\mu + \Delta & y^* \\
y & \mu - \Delta \\
\end{array}\right]
\label{matrix}
\ee
where the first and second rows correspond to $\nu$ and $\bar{\nu}$,
respectively.  We neglect the possibility of neutrino decay, so that
$M_\nu$ must be Hermitian, which implies that the mass parameters
$\mu$ and $\Delta$ (Dirac masses) are real.  Any nonvanishing $\Delta$
is a violation of the CPT constraint that a particle and its
antiparticle must have the same mass.  Any nonvanishing $y$ (Majorana
mass), which mixes $\nu$ and $\bar{\nu}$, is a violation of L
conservation.  (Note that limits on CPT violation entering via a heavy
Majorana mass have been considered in Ref.~\cite{mocioiu}).

The eigenstates of $M_\nu$ are
\be
|\nu_+\rangle = 
\cos\theta_{\nu\bar{\nu}}\,|\nu\rangle + 
e^{i\phi} \sin\theta_{\nu\bar{\nu}}\, |\bar{\nu}\rangle
\label{nu+}
\ee
with mass $m_+ = \mu + \sqrt{|y|^2 + \Delta^2}$, and
\be
|\nu_-\rangle =
-\sin\theta_{\nu\bar{\nu}}\,|\nu\rangle + 
e^{i\phi} \cos\theta_{\nu\bar{\nu}}\,|\bar{\nu}\rangle
\label{nu-}
\ee
with mass $m_- = \mu - \sqrt{|y|^2 + \Delta^2}$.  The neutrino-antineutrino
mixing angle is given by 
\be
\tan{2\theta_{\nu\bar{\nu}}} = \frac{|y|}{\Delta}\,,
\label{angle}
\ee
and $\phi = \arg(y)$.

Under CPT, $\nu$ and $\bar{\nu}$ transform as
\be
CPT |\nu\rangle = e^{i\xi} |\bar{\nu}\rangle,\;
CPT |\bar{\nu}\rangle = e^{i\xi} |\nu\rangle\,,
\ee
where we have left the phase $\xi$ free.  Since CPT is
an antiunitary operator,
\be
CPT (CPT |\nu\rangle) = CPT (e^{i\xi} |\bar{\nu}\rangle) =
e^{-i\xi} CPT |\bar{\nu}\rangle = e^{i(\xi - \xi)} |\nu\rangle = 
|\nu\rangle\,.
\ee
From Eqs.~(\ref{nu+},\ref{nu-}),
\be
CPT |\nu_+\rangle =
e^{i(\xi-\phi)} \left[
\sin\theta_{\nu\bar{\nu}}\,|\nu\rangle + 
e^{i\phi}\cos\theta_{\nu\bar{\nu}}\, |\bar{\nu}\rangle
\right]
\label{cptnu+}
\ee
and
\be
CPT |\nu_-\rangle =
-e^{i(\xi-\phi)} \left[
-\cos\theta_{\nu\bar{\nu}}\,|\nu\rangle + 
e^{i\phi}\sin\theta_{\nu\bar{\nu}}\, |\bar{\nu}\rangle
\right]\,.
\label{cptnu-}
\ee
Comparing with Eqs.~(\ref{nu+},\ref{nu-}), one sees that the mass
eigenstates $\nu_{\pm}$ are Majorana states (that is, CPT
self-conjugate apart from a phase) if and only if
$\theta_{\nu\bar{\nu}} = \pi/4$.  But, from Eq.~(\ref{angle}), this
value of $\theta_{\nu\bar{\nu}}$ corresponds to $\Delta = 0$; i.e., to
an absence of CPT violation.  If $\Delta \ne 0$, so that CPT is not
conserved, the neutrino mass eigenstates can no longer be Majorana
particles.  Nevertheless, if $y$ is nonzero, then there is $\nu -
\bar{\nu}$ mixing, L is not conserved, and $\beta\beta_{0\nu}$ can
occur.  But if it does occur, that would imply only that L is violated
and that there is a ``Majorana'' (i.e., $\nu - \bar{\nu}$ mixing) mass
term, and not that the neutrino mass eigenstates are CPT
self-conjugate.

It is interesting to compare this situation with the neutral kaon
system when a possible CPT-violating term is introduced as a
difference between the $K^\circ$ and $\bar{K}^\circ$ masses (e.g., see
Ref.~\cite{dib}).  To compare to neutrinos, we neglect kaon decay.
Then the kaon mass matrix is identical to the $M_\nu$ of
Eq.~(\ref{matrix}), but with the first and second rows now
corresponding to $K^\circ$ and $\bar{K}^\circ$, respectively.  The
neutral kaon mass eigenstates (these correspond to $K_S$ and $K_L$ in
the usual case) are described by Eqs.~(\ref{nu+}) and (\ref{nu-}),
with $\nu$ replaced by $K^\circ$ and $\bar{\nu}$ replaced by
$\bar{K}^\circ$.  The CPT conjugates of these states are described by
Eqs.~(\ref{cptnu+}) and (\ref{cptnu-}), with the same replacements.
Once again, the mass eigenstates are CPT self-conjugate if and only if
the CPT-violating parameter $\Delta$ vanishes.  Otherwise, each of
them differs from its CPT conjugate.  Nevertheless, so long as the
``Majorana'' $K^\circ - \bar{K}^\circ$ mixing term $y$ is present,
then strangeness (playing the role of L) is not conserved.  A kaon
born as a $K^\circ$ can evolve into a $K^\circ - \bar{K}^\circ$
mixture.  Observing, via the scattering of the kaon, that such
evolution had occurred, would imply that strangeness is not conserved,
and in particular, that $K^\circ$ and $ \bar{K}^\circ$ mix, but not
that the mass eigenstates are self-conjugate.


\section{Neutrinoless Double Beta Decay}

\begin{figure}
\centerline{\includegraphics[width=3in]{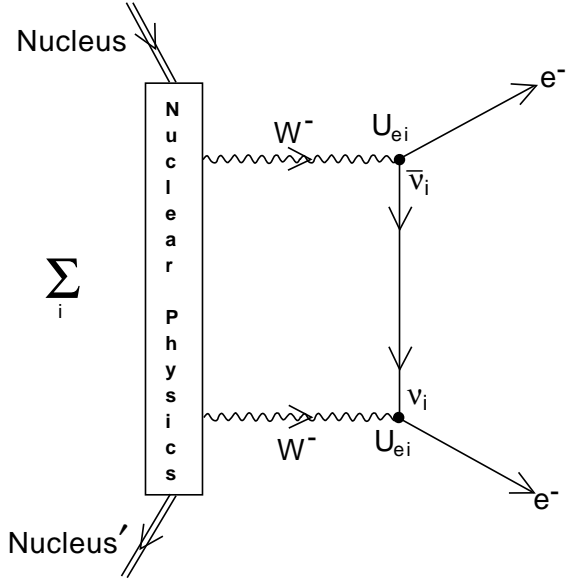}}
\caption{Diagram for neutrinoless double beta decay and the leptonic
amplitude considered below.  This diagram corresponds to the
CPT-conserving case; the modifications for the CPT-violating case are
considered below.}
\end{figure}

Neutrinoless double beta decay~\cite{2breviews} is the process
\bea
(A,Z) \longrightarrow (A, Z+2) + 2e^-
\eea
in which a nucleus decays into another nucleus with the emission of
two electron; see Fig.~1.  The emission of two $W$ bosons by the
initial nucleus is described by a nuclear matrix element, not
considered here.  The full amplitude for this process is proportional
to just the leptonic amplitude, which is
\be
A = \sum_{i,h} 
\langle e^- e^- | H_W | e^- \nu_i W^- \rangle
\langle e^- \nu_i W^- | H_W | W^- W^- \rangle\,,
\ee
where $i$ labels the neutrino mass eigenstates and $h$ their helicities.
Each mass eigenstate can be represented as a superposition of
neutrinos and antineutrinos, so that
\be
| \nu_i \rangle = 
\sum_{\alpha=e,\mu,\tau} U_{\alpha i} |\nu_\alpha\rangle +
\sum_{\alpha=e,\mu,\tau} \bar{U}^*_{\alpha i} |\bar{\nu}_\alpha\rangle\,,
\ee
which defines the matrices $U$ and $\bar{U}$ for neutrinos and
antineutrinos, respectively.  Absorbing a $\nu_l$ yields an $l^-$, and
similarly for a $\bar{\nu}_l$ and an $l^+$.  Because of the helicity
mismatch between the neutrino and antineutrino interaction, the leptonic 
amplitude is proportional to
\be
A \propto \sum_i m_i U_{ei} \bar{U}_{ei}\,,
\label{amp-cpt}
\ee
where the sum is over mass eigenstates.  In particular, for the one-flavor 
case discussed above,
\be
A \propto
\sin\theta_{\nu\bar{\nu}} \cos\theta_{\nu\bar{\nu}} (m_+ - m_-) 
= \sin{2\theta_{\nu\bar{\nu}}} \sqrt{|y|^2 + \Delta^2} = |y|\,.
\label{amp2}
\ee
As expected, the Majorana mass term $y$ contributes to the amplitude.
In fact, with just a single neutrino flavor, it is the only
contribution, i.e., the value of $\Delta$ is irrelevant. In the more
relevant case where more than one neutrino family is involved, there
are in general contributions depending on the CPT-violating mass term
$\Delta$, and hence the LSND scale of 1 eV.  We show this with an
explicit example below.  Even at this stage, it is easy to understand
that if the neutrino and antineutrino mass matrices are different, the
diagonalizing matrices must also be different.  The difference will be
reflected in the matrices $U$ and $\bar{U}$ in Eq.~(\ref{amp-cpt}).
The degree of difference depends on $\Delta$, the CPT-violating mass
term, and of course vanishes when $\Delta = 0$.

Note that the usual CPT-conserving amplitude is obtained with
$\bar{U} = U$ in Eq.~(\ref{amp-cpt}).  At this point two remarks are
in order. First, one should remember that in a CPT-violating world
$\bar{U} \neq U$ and second, Eq.~(\ref{amp-cpt}) is obtained without
depending on the usual Feynman rules by constructing the amplitudes
from quantum mechanics and therefore is quite safe regardless whether
our neutrinos conserve CPT or not.  It must be emphasized that the
same expression can also be obtained by using the method suggested in
\cite{cpt2}, i.e., by calculating the matrix elements as if they
belonged to a CPT-conserving neutrino in an artificial background of
matter.  In this case (as it happens in field theory at finite
temperature) the Feynman rules for the physical vertices are the same
as those in the CPT-conserving theory. The propagators, however, are
different.  For momenta much higher than the neutrino masses, the
propagator in matter can be written as an expansion in powers of the
CPT violating mass term, whose first term is the standard propagator.

Double beta decay will have no {\it manifest} CPT violation, in the
sense that the CPT-conjugated process involving two positrons will
have exactly the same rate.  This observation follows from the fact
that the corresponding amplitude for two positrons is simply the
complex conjugate of that for two electrons.  However, CPT violation
does have an experimentally observable consequence.  Namely, since in
the CPT-violating case the double beta decay amplitude involves two
independent matrices $U$ and $\bar{U}$, the decay rate can reach
values that are outside those allowed in the usual CPT-conserving
case.

In order to make this statement even more transparent, we first recall
the result for the CPT-conserving case.  The usual expression
(assuming only two flavors and degenerate masses) is
\bea
m_{\beta \beta} = m_0 \mid \cos^2\theta \,+\,
\sin^2\theta \, e^{2i \alpha } \mid
\eea
where $\theta$ is the angle involved in the solar neutrino solution
and $\alpha$ is a CP violating phase characteristic of Majorana
neutrinos which does not appear in oscillation-related phenomena.
Since SNO sees 1/3 of the expected $\nu_e$ flux (i.e., less than 1/2),
the mixing angle must be less than maximal (the best-fit value is
about $30^\circ$)~\cite{solar}.  Thus there is a {\it minimum} value
for $m_{\beta \beta}$, obtained when $\alpha = \pi/2$.  The range for
$m_{\beta \beta}$ is thus
\be
m_0 \cos(2 \theta_\odot) < m_{\beta \beta} < m_0\,.
\ee
Presently, $\cos(2 \theta_\odot) \simeq 0.4$, and the overall mass
scale can be very small.  In the more general three-flavor
CPT-conserving case, the expression for $m_{\beta \beta}$ is more
complicated, but it retains the feature of a minimum value (except
at a singular point)~\cite{mbb}.

When CPT is violated, the neutrino and antineutrino mixing matrices
can be different, and so can the neutrino and antineutrino masses.  If
MiniBooNE confirms LSND, then we know that a mass of about 1 eV exists
(this is the case of most interest to CPT violation).  This mass scale
is larger than required in the absence of the LSND oscillation signal,
raising hope that it might give a large neutrinoless double beta decay
signal.  In general, it is much more complicated to treat the full
problem, where there are now six mass eigenstates, each containing
components of both neutrinos and antineutrinos of all three flavors.
This can be visualized by merging the two spectra in Fig.~(1) in
Ref.~\cite{cpt1}, and allowing neutrino-antineutrino admixtures,
permitted by the Majorana masses.  That combined spectrum is
consistent with the solar, atmospheric, and LSND data.

However, we know that for this spectrum the state with mass
corresponding to the LSND mass scale is the dominant term in the sum
for $m_{\beta \beta}$ given by Eq.~(\ref{amp-cpt}), which is then
reduced to
\be
m_{\beta \beta} \simeq m_{LSND} U_{e,LSND} \bar{U}_{e,LSND}\,.
\ee
In this picture, $|U_{e,LSND}|^2$ represents the electron neutrino
content of the highest mass eigenstate, and similarly for
$|\bar{U}_{e,LSND}|^2$ and the electron antineutrino content.  In
order to explain all the data, the electron antineutrino content
should be very large ($\simeq 99\%$)~\cite{cpt2}, and the remainder
can be filled by other neutrino and antineutrino flavors.  These
proportions in particular can satisfy the bounds on reactor
$\bar{\nu}_e$ disappearance.  In the case where this remainder is
mostly the electron neutrino component, then $m_{\beta \beta}$ reaches
it maximal value, given by $\simeq 0.1 \, m_{LSND}$.  On the other
hand, there is no lower limit of the electron neutrino content, so
finally,
\be
0 < m_{\beta \beta} < 0.1 \, m_{LSND}\,.
\ee
Though beyond the scope of this work, it would be very interesting
to consider in more detail the allowed numerical range of 
$m_{\beta\beta}$ in realistic three-flavor models with CPT violation.


\section{Conclusions}

Present and forthcoming experiments devoted to studying neutrino
oscillations should give us much more information about the neutrino
mass-squared differences and mixing angles.  However, in order to
really understand the neutrino sector we will also need to know the
absolute mass scale and the nature (Dirac or Majorana) of the neutrino
mass terms.  Neutrinoless double beta decay experiments are crucial in
this respect, as well as for studying lepton flavor violation.

We have explored the nature of neutrinos when CPT is violated.
Contrary to the widespread belief that CPT-violating neutrinos can
have only a Dirac character and therefore no neutrinoless beta decay
can be expected, we have shown that CPT violation can also be seen in
neutrinoless double beta decay experiments.  As an important general
point, though Majorana neutrino masses are allowed, in general, there
are no longer Majorana neutrino states in the conventional sense.  If
the CPT-violating neutrino mixing model is chosen to explain the LSND
result, then at least one mass is of order the large LSND scale of 1
eV.  This can increase the neutrinoless double beta decay rate
relative to the usual case in which all neutrino masses can be small.
On the other hand, due to the freedom in the mixing between the
neutrino and antineutrino in each mass eigenstate, a greater degree of
suppression in the effective mass that appears in neutrinoless beta
decay is also possible.


\subsection*{Acknowledgements}
\noindent
We thank J. Lykken for discussions.  G.B., J.F.B (as the David
N. Schramm Fellow), and B.K. were supported by Fermilab, which is
operated by URA under DOE contract No.\ DE-AC02-76CH03000.  J.F.B.
was also supported by NASA under NAG5-10842.  L.B. was supported by
the Sloan Foundation.


\end{document}